\documentclass{article}
\usepackage[tcidvi]{graphicx}
\usepackage[english]{babel}
\usepackage{amsmath,amsfonts,graphics,color,colortbl,fontsmpl,amssymb,amsfonts,epsfig,rotating,float,amscd,multicol,layout}
\title{Negativity and Concurrence for two qutrits}
\author{Suranjana Rai$^\dagger$\quad and Jagdish R. Luthra\footnote{e-mail:
jluthra@uniandes.edu.co}\\
\small{($\dagger$) Raitech, Tuscaloosa, AL 35405}\\
\small{($*$) Departamento de F\'isica, Universidad de los Andes,
A.A. 4976 Bogot\'a, Colombia}}
\date{July 26, 2005}
\begin{document}
\maketitle
\begin{abstract}
Two measures of entanglement, negativity and concurrence are
studied for two qutrits. An operator origin of negativity is
presented and an analytic formula connecting the two measures is
derived.
\end{abstract}

Entanglement is well understood for two qubits. A good measure of
entanglement is the concurrence defined by Wootters
\cite{wootters}. It is related to the entanglement of formation
(EOF) which, for pure states, is the entropy of the reduced
density matrix. A great deal of work is going on to understand
both multipartite entanglement and entanglement of higher
dimensional systems. But a good measure or set of measures to
completely specify general qudit entanglement has not yet been
found. In this paper, we explore the connection between different
measures of entanglement for pure states of two qutrits, in
particular, the relation between negativity and concurrence.\\

Generalizing concurrence to higher-dimensional bipartite states is
not a mere extension of the two-qubit concurrence. A generalized
concurrence known as the I-concurrence has been obtained for
qudits\cite{rungta}. But I-concurrence is not a fully satisfactory
measure of entanglement since it is not a monotonic function of
the EOF. It does indeed reduce to the EOF for some special cases,
but not in general.\\

We look at another measure of entanglement, the negativity defined
by Vidal and Werner \cite{vidal}. Negativity is an entanglement
monotone so it does not change under local operations and
classical communications (LOCC). Vidal defines negativity of a
bipartite system described by the density matrix $\rho$, as
$N(\rho)$
\begin{eqnarray}
N(\rho) &=& \frac{||\rho^{T_A}||_1 - 1}{2}\
\end{eqnarray}
where $\rho^{T_A}$ is the partial transpose with respect to system
A, $||...||_1$ denotes the trace norm. Negativity is a
quantitative measure of the partial positive transpose (PPT), the
Peres criteria \cite{peres}. It measures how negative the
eigenvalues of the density matrix are after the partial transpose
is taken. It is the absolute value of the sum of the negative
eigenvalues of the partially transposed density matrix. According
to the Peres criterion of separability, density matrices with NPT
are entangled \cite{peres}. The Peres criterion is necessary and
sufficient for qubit-qubit and qubit-qutrit systems. For systems
of higher dimensions, it is necessary but not sufficient, since,
there are states with PPT which are entangled. These states are
said to have bound entanglement since entanglement can not be
distilled from these states.\\

The significance of negativity can be seen from the quantity
logarithmic negativity, which Vidal and Werner go on to define as

\begin{eqnarray}
E_N(\rho) &=& log_2||\rho^T_A||_1.
\end{eqnarray}
$E_N$ is an upper bound on the entanglement of distillation(EOD),
which for pure states reduces to the von Neumann entropy. But
$E_N(\rho)\geq E(\rho)$, so negativity gives a larger value than
the entropy and the equality holds for pure maximally entangled
states.\

Lee et. al., generalize the negativity to higher dimensions
\cite{Lee}

\begin{eqnarray}
N(\rho) &=& \frac{||\rho^{T_A}||_1 -1}{d-1}
\end{eqnarray}

where $d$ is smaller of the dimension of the bipartite system. We
use this negativity and it ranges from zero to one.\\

A general bipartite pure state for qutrits in the Schmidt form can
be written as
\begin{eqnarray}
|\psi\rangle &=& \sum_{i=1}^d k_i|ii\rangle
\end{eqnarray}
with normalization condition
\begin{eqnarray}
k_1^2 +k_2^2 +k_3^2 &=& 1
\end{eqnarray}
where $k_1,k_2$ and $k_3$ are the Schmidt coefficients, which are
non-negative real numbers.

The negativity $N$ of this state $|\psi\rangle$ is given in terms
of Schmidt coefficients as

\begin{eqnarray}
N(\rho) &=& \sum_{i<j} k_i k_j
\end{eqnarray}

For the two qutrit pure state, the negativity has the explicit form
\begin{eqnarray}
N(\rho) &=& (k_1k_2 + k_2k_3 + k_3k_1)
\end{eqnarray}

We also show that the negativity can be obtained by the action of
a ladder operator\cite{jay}, $X = X_1\otimes X_2$, where $X_1$
acts on the first qutrit and $X_2$ acts on the second qutrit. The
operator $X$ is defined as $X|11\rangle = |22\rangle , X|22\rangle
= |33\rangle$ and  $X|33\rangle =|11\rangle$.

In general
\begin{eqnarray}
X|i,i\rangle =|i+1,i+1\rangle, \text{ mod(d)}
\end{eqnarray}

The operator $X$ transforms the state $|\psi\rangle$ into a shifted
state by the ladder action. Just as Wootters concurrence is defined
using the Pauli spin matrix $\sigma_y$ as a spin flip operator, we
define the action of the ladder operator $X$, on a state and take
the inner product with the original state. This is similar in spirit
to the origin of concurrence for qubits. But it turns out to be not
the concurrence as in the qubit case but the Vidal negativity. The
operator $X$ is analogous to the flip operator $\sigma_y$. It is
easy to see that negativity can be obtained as the expectation value
of the $X$ operator in the state $|\psi\rangle$.
\begin{eqnarray}
N(\rho) =\langle X \rangle =(k_1k_2 + k_2k_3 + k_3k_1)
\end{eqnarray}
This expression is the same as Eq.(7). This equation also holds in
general for qudits
\begin{eqnarray}
N(\rho) =\langle X \rangle =\sum_{i<j} k_i k_j
\end{eqnarray}
We now connect the concurrence to the negativity. Cereceda
\cite{cereceda} has obtained a result for the concurrence
\cite{cereceda} $C$ of two qutrits  on the lines of the
I-concurrence \cite{rungta}.
\begin{eqnarray}
C(|\psi\rangle &=& \sqrt{3(k_1^2k_2^2 + k_2^2k_3^2 + k_3^2k_1^2)}
\end{eqnarray}
The negativity and concurrence provide two important measures of
entanglement. We are interested in exploring the relation between
negativity $N$ and concurrence $C$ and the conditions under which
the two are equal. Perhaps this connection might be useful for
finding the various quantities that are necessary for a complete
description of the entanglement. There exists a simple algebraic
relation between N and C for two qutrits
\begin{eqnarray}
N^2 &=& \frac{C^2}{3} \pm 2(k_1k_2k_3)\sqrt{{1+2N}}
\end{eqnarray}

Concurrence for two qubits is a monotonic function of the EOF, so
it is a good measure of entanglement. However, concurrence in
three dimensions is no longer a monotonic function of the EOF
\cite{cereceda}. For qubits, concurrence uses conjugation. The
conjugation is done by the Pauli operator $\sigma_y$. In higher
dimensions, conjugation is not so simple. There is no one
conjugated state. However, there is the inverted state which is
the sum of all the states available to the system. A pure state
after inversion goes in general to a mixed state. The concurrence
generated in this way has been called $I$-concurrence
\cite{rungta}, as distinct from the $\Theta$-concurrence
\cite{uhlmann}, which is based on antilinear conjugation. For
qutrits and higher dimensional systems, antilinear conjugation
fails as transformation matrices are singular.\

Both concurrence and negativity arise as a sum of entanglement
between pairs of levels. Concurrence is the root mean square of
pair-wise products of the Schmidt coefficients while negativity is
the sum of pair-wise products of the coefficients. The difference
contains a part which occurs with the product of the three Schmidt
coefficients, $k_1k_2k_3$. This product vanishes when three-level
entanglement is absent,i.e., when one of the Schmidt coefficient
is zero, in which case the N and C differ from each other by just
a scaling factor and now concurrence becomes a monotonic function
of the EOF. That this difference is in terms of the explicit
three-way entanglement of the two qutrits is possibly of some
significance.\

The negativity is upper bounded by the Cereceda concurrence. It is
interesting to note that $C\geq N$. The maximum value of $C$ and
$N$ are one for maximally entangled states.\

\begin{figure}[h!]
\centering \epsfig{file=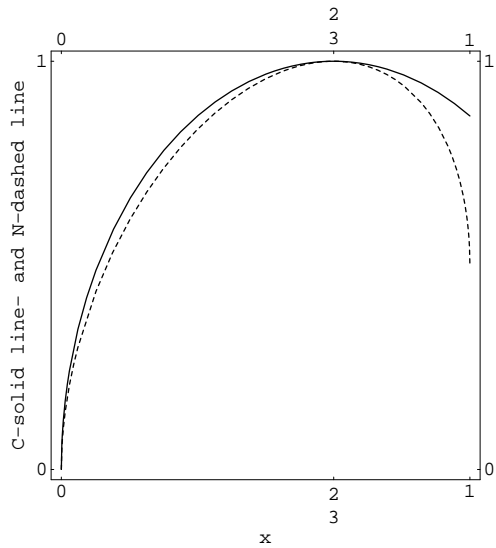,width=.6\linewidth}
\caption{Concurrence $C$ and negativity $N$ for two qutrits in
pure state. Concurrence $C$ is shown by the solid upper line, and
negativity $N$ by the lower dashed line, against the parameter
$x$, with $k_1=k_2=\sqrt{x/2}$ and $k_3=\sqrt{(1-x)}$. See also
Figs. \ref{C3D} and \ref{N3D} for a spherical parametrization of
$C$ and $N$, respectively.}\label{CNpure}
\end{figure}

Fu et. al. \cite{fu}, have studied the violation of
Clauser-Horne-Shimony-Holt (CHSH) for two qutrits. The degree of
entanglement $P_E$ they obtain is related to negativity as $P_E =
N$. Cereceda has compared his concurrence $C$ with $P_E$ for
states of two qutrits. We connect N with C for states with $k_1 =
k_2 = \sqrt{x/2}$ and $k_3 =\sqrt{(1-x)}$. This is shown in
Fig.\ref{CNpure}.

\begin{figure}[h!]
\centering \epsfig{file=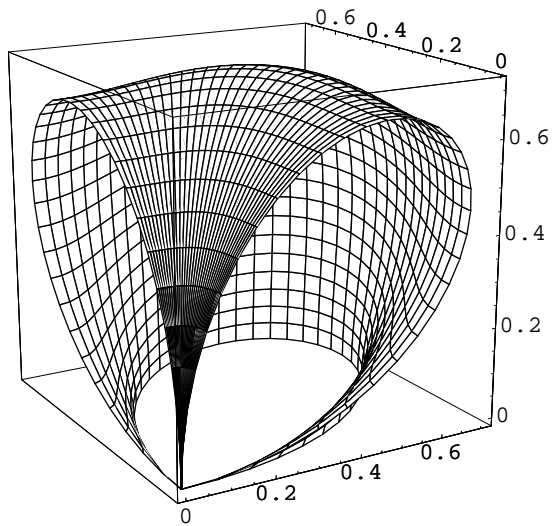,width=.6\linewidth}
\caption{Concurrence $C$ of two qutrits in a pure state, in
spherical coordinates $k_1=\sin\theta\cos\phi$,
$k_2=\sin\theta\sin\phi$, and $k_3=\cos\theta$, for
$0\leq\theta\leq\frac{\pi}{2}$ and
$0\leq\phi\leq\frac{\pi}{2}$.}\label{C3D}
\end{figure}

\begin{figure}[h!]
\centering \epsfig{file=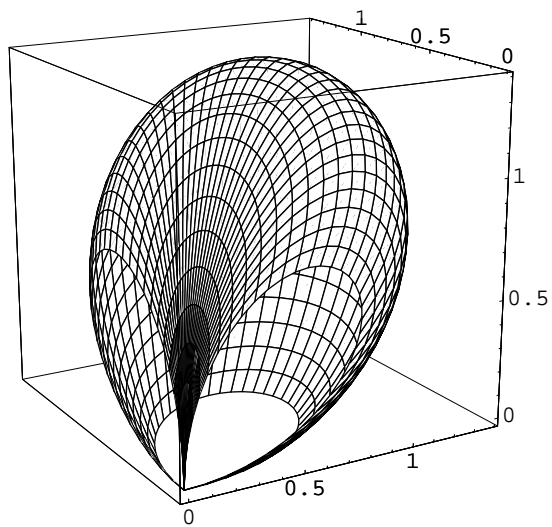,width=.6\linewidth}
\caption{Negativity $N$ of two qutrits in a pure state, in
spherical coordinates $k_1=\sin\theta\cos\phi$,
$k_2=\sin\theta\sin\phi$, and $k_3=\cos\theta$, for
$0\leq\theta\leq\frac{\pi}{2}$ and
$0\leq\phi\leq\frac{\pi}{2}$.}\label{N3D}
\end{figure}
\vspace{1cm}When one of the $k's$ is zero, the two qutrit state
reduces to a two-qubit state. In this case, even though $C$ goes
from zero to $\sqrt{3}/2$, the $N$ goes from zero to $1/2$.\\

In Fig. 2 and 3. we show the three-dimensional plots of $N$ and
$C$ using a general three parameter two qutrit state. The maximum
values of $N$ and $C$ are along the body diagonals of the cubes.
The variation of $C$ is flatter while the $N$ has a more well
defined lobe around its maximum direction.\\

In conclusion, $N$ and $C$ provide two useful quantities for
describing the entanglement property and are found to be related.
In higher dimensions, such a connection is more complicated but
may be fruitful to study.\\

We thank Cesar Herreno-Fierro for help with graphs and useful
discussions.

\end{document}